\documentclass[fleqn,usenatbib]{mnras}
\usepackage{newtxtext,newtxmath}
\usepackage[T1]{fontenc}
\usepackage{ae,aecompl}
\usepackage{graphicx}	
\usepackage{amsmath}	
\usepackage{amssymb}
\usepackage{times}
\usepackage{verbatim}

\title[Fundamental Plane evolution in IllustrisTNG]{Redshift Evolution of the Fundamental Plane Relation in the IllustrisTNG Simulation}
\author[S. Lu et al.]
{Shengdong Lu$^{1,2}$\thanks{E-mail: \url{lushengdong@nao.cas.cn}},
Dandan Xu$^{3}$,
Yunchong Wang$^{4,3}$,
Shude Mao$^{3,1}$,
Junqiang Ge$^{1}$,
\and
Volker Springel$^{5}$,
Yuan Wang$^{6}$,
Mark Vogelsberger$^{7}$,
Jill Naiman$^{8,9}$,
Lars Hernquist$^{8}$
\\
$^{1}$National Astronomical Observatories, Chinese Academy of Sciences, 20A Datun Road, Chaoyang District, Beijing 100101, China\\
$^{2}$University of Chinese Academy of Sciences, Beijing 100049, China\\
$^{3}$Department of Astronomy and Tsinghua Center for Astrophysics, Tsinghua University, Beijing 100084, China\\
$^{4}$Department of Physics, Tsinghua University, Beijing, 100084, China\\
$^{5}$Max-Planck-Institut f\"{u}r Astrophysik, Karl-Schwarzschild-Str. 1, D-85748, Garching, Germany\\
$^{6}$School of Physics, Nankai University, Tianjin 300071, China\\
$^{7}$Kavli Institute for Astrophysics and Space Research, Department of Physics, MIT, Cambridge, MA 02139, USA\\
$^{8}$Harvard-Smithsonian Center for Astrophysics, 60 Garden Street, Cambridge, MA, 02138, USA\\
$^{9}$The School of Information Sciences at the University of Illinois Urbana-Champaign, 501 E Daniel St, Champaign, IL 61820, USA
}

\date{Accepted ***. Received ***; in original form ***}

\pubyear{2020}

\begin{document}
\label{firstpage}
\pagerange{\pageref{firstpage}--\pageref{lastpage}}
\maketitle
\begin{abstract}
We investigate the fundamental plane (FP) evolution of early-type galaxies in the IllustrisTNG-100 simulation (TNG100) from redshift $z=0$ to $z=2$. We find that a tight plane relation already exists as early as $z=2$. Its scatter stays as low as $\sim 0.08$ dex across this redshift range. Both slope parameters $b$ and $c$ (where $R \propto \sigma^b I^c$ with $R$, $\sigma$, and $I$ being the typical size, velocity dispersion, and surface brightness) of the plane evolve mildly since $z=2$, roughly consistent with observations. The FP residual $\rm Res$ ($\equiv\,a\,+\,b\log \sigma\,+\,c\log I\,-\,\log R$, where $a$ is the zero point of the FP) is found to strongly correlate with stellar age, indicating that stellar age can be used as a crucial fourth parameter of the FP. However, we find that $4c+b+2=\delta$, where $\delta \sim 0.8$ for FPs in TNG, rather than zero as is typically inferred from observations. This implies that a tight power-law relation between the dynamical mass-to-light ratio $M_{\rm dyn}/L$ and the dynamical mass $M_{\rm dyn}$ (where $M_{\rm dyn}\equiv 5\sigma^2R/G$, with $G$ being the gravitational constant) is not present in the TNG100 simulation. Recovering such a relation requires proper mixing between dark matter and baryons, as well as star formation occurring with correct efficiencies at the right mass scales. This represents a powerful constraint on the numerical models, which has to be satisfied in future hydrodynamical simulations.
\end{abstract}
%for the FPs in TNG we find that 4c+b+2=\delta where \delta \simeq 1, rather than zero as is typically inferred from observations. This may imply that..

\section{Introduction}
\label{sec:introduction}
Early-type galaxies (ETGs) are the final products of the hierarchical assembly of galaxies via mergers and accretion \citep[e.g.,][]{Toomre_and_Toomre(1972),Bender_et_al.(1992)}. They are found to obey several empirical scaling relations, e.g., the Faber-Jackson relation \citep{Faber_and_Jackson(1976)} between a galaxy's velocity dispersion $\sigma$ and its luminosity $L$, the relation between the supermassive black hole mass $M_{\rm BH}$ and its host galaxy bulge's luminosity $L$ or velocity dispersion $\sigma$ \citep{Kormendy_and_Richstone(1995),Magorrian_et_al.(1998),Ferrarese_and_Merritt(2000),Husemann_et_al.(2016),Subramanian_et_al.(2016)}, as well as the famous Fundamental Plane (FP) relation \citep{Djorgovski_and_Davis(1987),Dressler_et_al.(1987),Jorgensen_et_al.(1996),Cappellari_et_al.(2006),La_Barbera_et_al.(2008)} that exists between galaxy size $R$, velocity dispersion $\sigma$, and surface brightness $I\equiv{L/(\pi R^2)}$. 

In particular, the tight FP relation reflects correlations among a galaxy's structural properties, dynamics, and star-formation activities. The FP can be expressed as:
\begin{equation}
\label{eq:FP}
\log R = a + b \log \sigma + c \log I,
\end{equation}
where $a$, $b$ and $c$ are the plane variables. The origin of the FP can be understood from virial theorem \citep{Faber_et_al.(1987)}, which links a galaxy's gravitational potential energy $-GM_{\rm tot}^2/R$ with its kinetic energy $T_{\rm kin}\propto M_{\rm tot} \sigma^2 /2$, where $M_{\rm tot}$ is the total mass of a galaxy; $R$ and $\sigma$ are characteristic size and velocity dispersion measurements. Rewriting $M_{\rm tot}$ in terms of luminosity $L$ and a total mass-to-light ratio $M_{\rm tot}/L$, the virial theorem can be translated into a plane relation:
\begin{equation}
\label{eq:virial}
\log R = K_{\rm SD} + 2 \log \sigma -  \log I  - \log M_{\rm tot}/L ,
\end{equation}
where $K_{\rm SD}$ is a normalization factor depending on a galaxy's structural properties and dynamics (see \citealt{Taranu_et_al.(2015)} for a more detailed discussion on this topic). The observed FP relation typically has parameters $b\neq 2$ and $c\neq -1$, which may not necessarily mean the breakdown of the virial theorem. It also manifests a breakdown of homologies in galactic structural and dynamical distributions, i.e., constant $K_{\rm SD}$ \citep[e.g.,][]{Ciotti_et_al.(1996),Pahre_et_al.(1998a),Pahre_et_al.(1998b),Bertin_et_al.(2002),Trujillo_et_al.(2004),Saglia_et_al.(2010)} as well as of constant mass-to-light ratios $M_{\rm tot}/L$ \citep[e.g.,][also see references below]{Pahre_et_al.(1995),Cappellari(2016)}. Non-homologies and non-constant $M_{\rm tot}/L$ ratios can be caused by, for example, variations in the density profiles \citep[e.g.,][]{Schombert(1986)}, in the velocity dispersion anisotropies \citep[e.g.,][]{Davies_et_al.(1983),Busarello_et_al.(1992),Ciotti_et_al.(1996),Nipoti_et_al.(2002)}, in the different spatial distributions of dark matter and visible matter \citep[e.g.,][]{Ciotti_et_al.(1996),Humphrey_and_Buote(2010)}, as well as in the stellar initial mass function (IMF), resulting in different stellar populations \citep[e.g.,][]{Renzini_and_Ciotti(1993)}. Such non-homologies have been observed in galaxies of different masses, with different formation histories, and in different environments \citep[e.g.,][]{Pahre_et_al.(1998b),van_Dokkum_et_al.(2001),Bernardi_et_al.(2003),Lanzoni_et_al.(2003),Bernardi_et_al.(2006),Saglia_et_al.(2010)}. 

The tightness of the {\it observed} FP is even stronger when the relation is translated into the so-called $\kappa$-space \citep{Bender_et_al.(1992)} via an orthogonal transformation. By assuming a so-called `dynamical mass' $M$ ($\propto \sigma^2 R$), this edge-on view can be interpreted as the FP projected onto the $\log M/L - \log M$ plane, with the dynamical mass taking the mathematical form:
\begin{equation}
\log M = A + 2 \log \sigma + \log R,
\label{eq:DynM}
\end{equation}
where $A$ is a normalization factor (e.g., \citealt{Faber_et_al.(1987)}; \citealt{van_Albada_et_al.(1995)}). By invoking an exact power-law relation between $M/L$ and $M$, i.e., $M/L \propto M^\alpha$, where $\alpha>0$ reflects that more massive galaxies have larger mass-to-light ratios (also referred to as the so-called `tilt of the FP\footnote{Note that homologies in $K_{\rm SD}$ and $M/L$, i.e., plane parameters of $b = 2$ and $c = -1$ correspond to $\alpha=0$, i.e., $M/L$ being independent of galaxy mass $M$.}'), we have another alternative relation after rewriting $M$ and $L$ in terms of $R$, $\sigma$, and $I$:
\begin{equation}
\log R = A^{\prime}+\frac{2\left(1-\alpha\right)}{1+\alpha}\log \sigma+\frac{-1}{1+\alpha}\log I,
\label{eq:fp2}
\end{equation}
where $A^{\prime}$ is a normalization factor. Comparing this with Eq.(\ref{eq:FP}), we find:
\begin{equation}
\left\{
\begin{aligned}
&\frac{2\left(1-\alpha\right)}{1+\alpha}=b,\\
&\frac{-1}{1+\alpha}=c.
\end{aligned}
\right.
\label{eq:fp3}
\end{equation}
Thus, to derive the power-law relation between $M/L$ and $M$, we must have:
\begin{equation}
\left\{
\begin{aligned}
&4c+b+2=0,\\
&\alpha=1+\frac{b}{2c}.
\end{aligned}
\right.
\label{eq:4cb2}
\end{equation}
Observed early-type galaxies broadly satisfy these relations at various redshifts extending to $z>1$ \citep[e.g.,][]{Bender_et_al.(1992),Jorgensen_et_al.(1996),Jorgensen_et_al.(2006),Hyde_and_Bernardi(2009), Cappellari_et_al.(2013a)}. Many studies have tried to understand why such a tight correlation between the dynamical mass and the dynamical mass-to-light ratio exists for early-type galaxies, but it is not yet fully understood \citep[e.g.,][]{Renzini_and_Ciotti(1993),Ciotti_et_al.(1996),Graham_and_Colless(1997),Pahre_et_al.(1998a)}.

One of the many utilities of the FP relation that is directly relevant to galaxy formation studies, is to use the offset of the plane's intercept $a$ from its local ($z=0$) zero point value to measure the redshift evolution of the mass-to-light ratio $M/L$, from which halo assembly histories (through the evolution of the galaxy mass) as well as star-formation histories (through the age of the stellar population) can be further inferred. This has been routinely carried out for observed early-type galaxies \citep[e.g.,][]{van_Dokkum_and_Franx(1996),Bender_et_al.(1998),van_de_Ven_et_al.(2003),Jorgensen_et_al.(2006)}. For this to work out, a couple of assumptions need to be made in practice. It is easy to show that under Eq.~(\ref{eq:DynM}), the corresponding dynamical mass-to-light ratio $M/L$ can be expressed in terms of FP variables:
\begin{equation}
\log M/L = (2c+b)/c \log \sigma - (1+c)/c \log R + [a/c-\log \pi + A].
\end{equation}
The change of the mass-to-light ratio $\Delta \log M/L$ over cosmic times can then be written as:
\begin{equation}
\Delta \log M/L = \Delta [(2c+b)/c \log \sigma] + \Delta [-(1+c)/c \log R] + \Delta [a/c].
\label{eq:DeltaM2Lg}
\end{equation}
As can be seen, the redshift evolution of $M/L$ can result from galaxy structural and dynamical evolution (due to $\Delta R$ and $\Delta \sigma$), from the `rotation' of the FP (due to $\Delta b$ and $\Delta c$), and from changes in the plane's zero point (due to $\Delta a$) over cosmic time. 

Suppose we are allowed to assume that the FP slope parameters $b$ and $c$ change little within a given redshift range, then Eq.~(\ref{eq:DeltaM2Lg}) reduces to:
\begin{equation}
\Delta \log M/L = (2c+b)/c \Delta \log \sigma -(1+c)/c \Delta \log R + (\Delta a)/c.
\label{eq:DeltaM2Lf}
\end{equation}
Further assuming that (\textbf{i}) the dynamical mass of an early-type galaxy does not change significantly within a given redshift range, i.e., $\Delta \log M = 0$ thus $\Delta \log R = -2 \Delta \log \sigma$, plus that Eq.~(\ref{eq:4cb2}) stands; or alternatively (\textbf{ii}) the velocity dispersion and size of the galaxy does not vary much within a given redshift range, i.e., $\Delta \log \sigma=0$ and $\Delta \log R=0$, then Eq.~(\ref{eq:DeltaM2Lf}) reduces to:
\begin{equation}
\Delta \log M/L = (\Delta a)/ c.
\label{eq:DeltaM2La}
\end{equation}

As can be seen, when all the involved assumptions approximately hold, it is possible to estimate the evolution of $M/L$ by a simple approximation using the FP zero point offset in combination with the local FP parameters, as indicated by Eq.~(\ref{eq:DeltaM2La}). Observed early-type galaxies seem to more or less obey some of the key assumptions listed above, but not necessarily all. For example, the FP slopes $b$ and $c$ are observed to have experienced some moderate evolution with cosmic time \citep[e.g.,][]{di_Serego_Alighieri_et_al.(2005), Treu_et_al.(2005), Jorgensen_et_al.(2006), Saglia_et_al.(2010)}. Both morphological (size and shape) and dynamical evolution of early-type galaxies have been theoretically proposed \citep[e.g.,][]{Biermann_and_Shapiro(1979),Kobayashi(2005),Khochfar_and_Silk(2006),Fan_et_al.(2008),Hopkins_et_al.(2009),Genel_et_al.(2018)}, which are supported by observational evidence \citep[e.g.,][]{van_Dokkum_et_al.(2008),Saglia_et_al.(2010),van_der_Wel_et_al.(2011),Chevance_et_al.(2012)} and may result from various internal feedback processes as well as galaxy mergers. One should keep in mind, however, that any strong violation of the assumptions above could result in the derived luminosity evolution (through Eq.~\ref{eq:DeltaM2La}) differing from the actual redshift evolution and thus introducing uncertainties to the estimation. Furthermore, in reality, observational samples can be even more complicated due to various selection effects \citep[e.g.,][]{van_der_Wel_et_al.(2005)} and progenitor bias \citep[e.g.,][]{van_Dokkum_et_al.(2001),Valentinuzzi_et_al.(2010),Saglia_et_al.(2010)}.

Accurate estimates of a FP's zero point are closely related to another interesting feature (or the fourth parameter) of the plane, i.e., the plane's scatter. Observed FPs, although tight, are found to possess some intrinsic scatter that cannot be explained purely by observational errors (see references listed above). The cause of this scatter has been attributed to variations in the stellar populations (\citealt{Gregg(1992),Prugniel_and_Simien(1996)}; \citealt{Jorgensen_et_al.(2006)}), in particular, their (post-merger) ages since the last major episode of star-formation (which are possibly triggered by  gas-rich merger events) in these early-type galaxies \citep{Forbes_et_al.(1998),Terlevich_and_Forbes(2002)}. This in fact is also consistent with the theoretical expectation for early-type galaxies with a given dynamical mass at a fixed redshift: changes in luminosity among galaxies, as reflected by the variation of the plane's zero point (see Eq.~\ref{eq:DeltaM2La}), result from the aging (passive evolution) of the stellar populations since their last major star-forming phase.

With the advancement of state-of-the-art numerical tools and collective efforts to simulate cosmic structure formation, a number of new-generation hydrodynamic cosmological simulations have been performed, which have significantly increased our abilities to understand galaxy formation within the current cosmology framework dominated by dark matter and dark energy. Projects of this type include the Illustris Simulations \citep{Genel_et_al.(2014),Vogelsberger_et_al.(2014a),Vogelsberger_et_al.(2014b),Nelson_et_al.(2015)}, the EAGLE Simulations \citep{Crain_et_al.(2015),Schaye_et_al.(2015)}, the Horizon-AGN project \citep{Dubois_et_al.(2014)}, as well as the latest IllustrisTNG simulation\footnote{\url{http://www.tng-project.org}} \citep{Marinacci_et_al.(2018),Naiman_et_al.(2018),Nelson_et_al.(2018),Pillepich_et_al.(2018b),Springel_et_al.(2018)} etc. These advanced cosmological simulations have provided us with large populations of galaxy samples, which agree with available observations to different degrees. Through detailed comparison analyses, one can not only calibrate the modelling parameters and guide the interpretation of observations, but also identify improper model treatment of the sub-grid physics in these simulations. \citet{Rosito_et_al.(2019)} and \citet{D'Onofrio_et_al.(2019)} investigated the 2D scaling relations in the EAGLE and Illustris simulations, respectively. They found that the observed 2D scaling relations are approximately reproduced by these two simulations. However, these relations show large scatters when projected from the 3D parameter space to 2D. Thus, we directly investigate the scaling relation in the 3D parameter space (the fundamental plane) in this work. 

The goal of this paper is multi-fold: (\textbf{i}) to examine whether the IllustrisTNG-100 simulation has produced a FP plane relation among the size, velocity dispersion, and the surface brightness of early-type galaxies; if so, then (\textbf{ii}) to establish how early such a plane relation emerges and how the plane parameters evolve with cosmic time; and most importantly, (\textbf{iii}) to investigate whether this plane is plausible compared to observations, i.e., whether Eq.~(\ref{eq:4cb2}) is met or equivalently, whether a power-law relation between $M/L$ and $M$ has been well produced by the simulation. 

The paper is organized as follows. In Section~\ref{sec:methodology}, we introduce the simulation (Section~\ref{sec:simulation}), describe our mock galaxy sample and galaxy morphology classification (Section~\ref{sec:samples}), and specify the FP fitting method (Section~\ref{sec:fpmethod}). Section~\ref{sec:result1} is devoted to presenting the FP slopes $b$ and $c$ and their redshift evolution. The redshift evolution of the FP scatter, zero point, and the relation between the FP scatter and stellar age are presented in Section~\ref{sec:result2}. The relation between the dynamical mass and the dynamical mass-to-light ratio of the TNG100 early-type galaxies is presented in Section~\ref{sec:result3}. Finally, we summarize our findings in Section~\ref{sec:conclusion}.

\section{Methodology}
\label{sec:methodology}
\subsection{The IllustrisTNG Simulations}
\label{sec:simulation}
\textit{The Next Generation Illustris Simulations} (IllustrisTNG) \citep{Marinacci_et_al.(2018),Naiman_et_al.(2018),Nelson_et_al.(2018),Pillepich_et_al.(2018b),Springel_et_al.(2018)} are a suite of state-of-the-art magneto-hydrodynamic cosmological galaxy formation simulations carried out in large cosmological volumes with the moving-mesh code \textsc{arepo} \citep{Springel(2010)}. The IllustrisTNG Simulations are built and improved upon the original Illustris Simulations with the same initial conditions \citep{Genel_et_al.(2014),Vogelsberger_et_al.(2013),Vogelsberger_et_al.(2014a),Vogelsberger_et_al.(2014b),Nelson_et_al.(2015)} but differ in the updated version of the galaxy formation model, including the addition of ideal magneto-hydrodynamics, a new active galactic nucleus (AGN) feedback model that operates at low accretion rates \citep{Weinberger_et_al.(2017)} and various modifications to the galactic winds, stellar evolution, and chemical enrichment schemes \citep{Pillepich_et_al.(2018a)}. In this work, we use its full-physics version with a cubic box of $110.7\,\mathrm{Mpc}$ side length (TNG100); it has been made publicly available\footnote{\url{http://www.tng-project.org/data/}} \citep{Nelson_et_al.(2019)}.  The mass resolutions of the TNG100-full physics version for baryonic and dark matter are $m_{\rm baryon}=1.4\times10^6\,{\rm M_{\odot}}$ and $m_{\rm DM}=7.5\times10^6\,{\rm M_{\odot}}$, with a gravitational softening length of $\epsilon = 0.74\,\mathrm{kpc}$. Gas cells are resolved in a fully adaptive manner with a minimum softening length of $0.19$ comoving $\mathrm{kpc}$. Galaxies in their host dark matter halos are identified using the {\sc subfind} algorithm \citep{Springel_et_al.(2001),Dolag_et_al.(2009)}.

\subsection{Early-type galaxy classification and sample selection}
\label{sec:samples}
We describe our criteria of sample selection and galaxy morphology classification in this section. To start with, we only select central galaxies to compose our sample, excluding all satellite galaxies whose host dark matter subhalos are identified by \textsc{subfind}. Each stellar particle is treated as a coeval stellar population following the \citet{Chabrier(2003)} initial mass function (IMF). We derive the stellar luminosity using the stellar population synthesis (SPS) model \textsc{galaxev} \citep{Bruzual_and_Charlot(2003)}. A semi-analytical dust attenuation treatment is applied to the stellar light based on neutral hydrogen density and metallicity to mock observations (see \citealt{Xu_et_al.(2017)} for details). Furthermore, we compute the surface brightness profiles within elliptical isophotes determined by the projected luminosity-weighted second moments of the galaxy, and perform a 1D S{\'e}rsic profile fitting within 0.05 to 3 times $R_{\rm hsm}$ to obtain the S{\'e}rsic index.
To ensure that every galaxy in our sample is sufficiently resolved, we only consider galaxies and their associated host dark matter halos with stellar masses $M_{\ast,\rm 30kpc} \geqslant 5\times10^9\,\rm M_{\odot}$. $M_{\ast,\rm 30kpc}$ is the total stellar mass within a radius of 30 kpc from the galaxy center, for which we adopt the position of the particle with the lowest gravitational potential in its host halo. 

Galaxy types are subject to different possible classification criteria. One can define a sample of early-type galaxies based on their more compact morphologies, more dispersion-dominated kinematics, redder colors and/or lower star-formation rates. The sole usage of a single individual criterion does not necessarily yield consistent galaxy-type determination (e.g., \citealt{Bottrell_et_al.(2017)},  \citealt{Wang_et_al.(2019)}, \citealt{Donnari_et_al.(2019)}). Therefore, we combine the light profile measured in the rest-frame SDSS $r$-band \citep{Stoughton_et_al.(2002)} with the specific star-formation rate (sSFR) to classify typical central early-type galaxies for this study. Here we set up the two criteria for galaxy-type classification:
\begin{enumerate}
\item Following \citet{Xu_et_al.(2017)}, we fit both a single de Vaucouleurs profile \citep{de_Vaucouleurs(1948)} (which is often used to describe the light distribution of an elliptical galaxy) and a single exponential profile (which is often used to describe the light distribution of a disk galaxy) to a galaxy's radial surface brightness distribution. An early-type galaxy should have the de Vaucouleurs model fit its light distribution better than the exponential profile.

\item It is a common practice in observations to separate early- and late-type galaxies with a certain specific SFR (sSFR) threshold \citep{McGee_et_al.(2011),Wetzel_et_al.(2013),Lin_et_al.(2014),Jian_et_al.(2018)}, which is also used in simulations \citep{Genel_et_al.(2018)}. Thus, we follow the practice of \citet{Genel_et_al.(2018)} and classify quenched galaxies (early-type galaxies) by quantifying their distance from the ridge of star-forming main-sequence galaxies. The ridge of main-sequence galaxies is defined as the mean specific SFR (sSFR) of galaxies with $M_{\ast,\rm 30kpc}<10^{10.5}\,\mathrm{M_{\odot}}$ at each selected redshift. Here we calculate the sSFR in an aperture with three-dimensional half stellar mass radius, $R_{\rm hsm}$. Galaxies are classified to be quenched if their sSFR is at least $1\,\rm dex$ below the ridge (for more details, see \citealt[Appendix A]{Genel_et_al.(2018)}). 
\end{enumerate}

\begin{figure*}
\centering
\includegraphics[width=1.9\columnwidth]{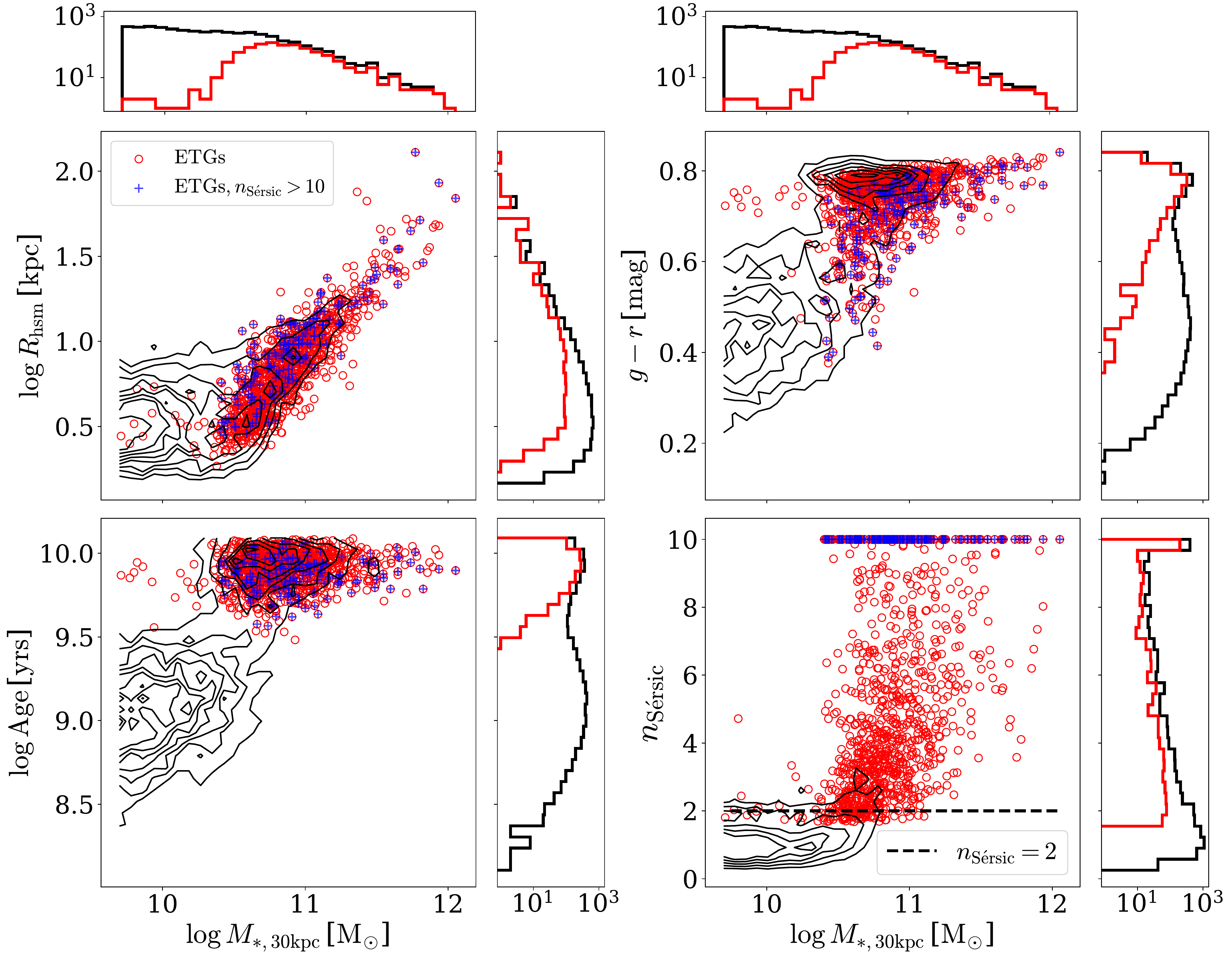}
\caption{Distributions of the three-dimensional half stellar mass radius $\log R_{\rm hsm}$, SDSS $g-r$ color, stellar age $\log \rm Age$, and S{\'e}rsic index $n_{\rm S\acute{e}rsic}$ versus $\log M_{\ast,\rm 30kpc}$ for selected early-type galaxies (red circles) and the full sample ($M_{\ast,\rm 30kpc} \geqslant 5\times10^9\,\rm M_{\odot}$) from TNG100 (black contours), respectively. The histograms are the projected distributions of the parameters for both the early-type galaxies (red) and the full sample (black) shown in each panel. The black dashed line in the lower right panel indicates $n_{\rm S\acute{e}rsic}=2$. The upper limit of $n_{\rm S\acute{e}rsic}$ is set to be 10 in the lower right panel and galaxies with $n_{\rm S\acute{e}rsic}>10$ are indicated with plus symbols at the centers of the circles in each panel.} 
\label{fig:pd}
\end{figure*}

If a galaxy satisfies both criteria above in {\it all} three principal projections (along X, Y and Z axes of the simulation box), it is then classified as an early-type galaxy. In this work, we view our selected galaxies in their X-projection. Fig.~\ref{fig:pd} shows the $z=0$ distributions of $\mathrm{log}\,R_{\rm hsm}$, SDSS $g-r$ color, luminosity-weighted stellar age within $R_{\rm hsm}$ ($\mathrm{log}\,\rm Age$), and S{\'e}rsic index ($n_{\rm S\acute{e}rsic}$) versus $\log M_{\ast,\rm 30kpc}$ (see \citealt{Xu_et_al.(2017)} for detailed descriptions of galaxy property extraction). As can be seen from the figure, the selected early-type galaxies indeed occupy the expected regions in the parameter space: they are typically more massive and bigger, redder and older, and have $n_{\rm S\acute{e}rsic}\gtrsim2$. Note that there are many old, red, and massive galaxies excluded from our samples. That is because they have small S{\'e}rsic index ($n_{\rm S\acute{e}rsic}<2$ or disky), even though they are close to ETGs in terms of star formation. Such galaxies have shown to be overproduced in IllustrisTNG (see \citealt[Fig. 8]{Rodriguez-Gomez_et_al.(2019)}), so we exclude them from our sample. In addition, there are galaxies with large $n_{\rm S\acute{e}rsic}$ ($>10$). They typically have power-law outer profiles and are not as cuspy as true $n_{\rm S\acute{e}rsic}>10$ profiles. This may be caused by several factors, such as mis-centering, gravitational softening, and the fact that we exclude the inner regions when we perform S{\'e}rsic profile fits (see \citealt{Xu_et_al.(2017)} for more details).

The above galaxy-type classification and sample selection criteria are applied to galaxies at the following redshifts, $z=[0.0,\,0.1,\,0.2,\,0.3,\,0.4,\,0.5,\,0.7,\,1.0,\,1.5,\,2.0]$, for which full snapshot data exist in the TNG100 simulation. This results in 1064, 944, 852, 745, 665, 606, 513, 365, 200, and 64 early-type galaxies selected at the above-mentioned redshifts, respectively. As the number of early-type galaxies within the required stellar mass range decreases significantly with redshift, the highest redshift in this study is limited to $z=2.0$ in order to maintain sizeable statistical samples and reliable galaxy morphologies. 

We also remind the reader that since the TNG100 simulation contains only a few galaxies which are in clusters, the majority of our sample galaxies are non-cluster early-type galaxies. Therefore we do not aim to address the environmental dependence of the FP relation in this work. We have also visually inspected the morphology of all selected galaxies at different redshifts. We confirm that galaxies that happen to be experiencing merger processes are very rare cases ($\sim 0.5\%$). Therefore our samples do not suffer from significant contamination by on-going merging systems. 

\subsection{Fundamental plane fitting}
\label{sec:fpmethod}
We adopt the following mathematical form to describe the FP relation:
\begin{equation}
\label{fpexpression}
\log \frac{R_{\rm hsm}}{\rm kpc}=a + b\,\log \frac{\sigma_{\mathrm{e}}}{\rm km\,s^{-1}} + c\,\log \frac{I_{\rm e}}{\mathrm{L_{\odot}\,kpc^{-2}}},
\end{equation}
where $\sigma_{\mathrm{e}}$ is the luminosity-weighted root-mean-square velocity of all stellar particles within $R_{\rm hsm}$ calculated  along a given line of sight (X-projection);
$I_{\rm e}$ is the rest-frame SDSS $r$-band mean surface brightness of all stellar particles within $R_{\rm hsm}$, defined as follows:
\begin{equation}
I_{\rm e} =  \frac{L}{\pi R_{\rm hsm}^2},
\end{equation}
where $L$ is the rest-frame SDSS $r$-band luminosity of the galaxy projected within $R_{\rm hsm}$.

We note that for simplicity, $R_{\rm hsm}$, which is directly measured from simulation data, is used here as an approximation of the commonly used 2D effective radius $R_{\rm eff}$ in observations. We use {\sc LTS\_PLANEFIT} described in \citet{Cappellari_et_al.(2013a)} to perform plane fitting on galaxies at redshifts shown in Table~\ref{table:table1}. {\sc LTS\_PLANEFIT} combines the Least Trimmed Squares robust technique of \citet{Rousseeuw_and_Van_Driessen(2006)} into a least-squares fitting algorithm and allows us to take observational error into account and exclude outliers. We also carry out the FP analysis using the aperture with 2D effective radius $R_{\rm eff}$ at several redshifts, which is the effective radius from the best-fit Multi-Gaussian Expansion (MGE)\footnote{The software is available from \url{https://www-astro.physics.ox.ac.uk/~mxc/software/}.} \citep{Emsellem_and_Monnet_and_Bacon(1994),Cappellari(2002)} formalism of the galaxy's surface brightness distribution, and verify that our main results remain unchanged.

To estimate the errors of the FP coefficients, we employ a bootstrap method. For each sample at a given reshift, we draw 500 bootstrap samples with sizes same as the original sample. Compared to the fitting error of the plane parameters fitted to the original sample, the standard deviations of the FP coefficients from bootstrapping are slightly larger. Thus, we take the standard deviation of the fitted FP coefficients from bootstrapping as the best-fit parameter uncertainties.

\section{Results}
\label{sec:results}
Table~\ref{table:table1} presents the best-fit plane's zero point $a$, the slope parameters $b$ and $c$, as well as the scatter $\Delta$ and the relative scatter $\Delta/\sigma_{\rm R}$ at various investigated redshifts, where $\Delta$ is the plane scatter and $\sigma_{\rm R}$ is the scatter in $\log R_{\rm hsm}$. $\Delta$ and $\sigma_{\rm R}$ are calculated as:
\begin{equation}
\left\{
\begin{aligned}
&\Delta = \sqrt{\frac{\sum_{i=1}^N \mathrm{Res}_i^2}{N}},\\
&\sigma_{\rm R} = \sqrt{\frac{\sum_{i=1}^N (\log R_{\mathrm{hsm},i}-\overline{\log R_{\rm hsm}})^2}{N}},
\end{aligned}
\right.
\label{eq:scatter}
\end{equation}
where $N$ is the total number of selected galaxies at each redshift; $\mathrm{Res}$ is the plane residual, which is calculated as $\mathrm{Res}\equiv a + b\log \sigma_{\mathrm{e}} + c \log I_{\rm e} - \log R_{\rm hsm}$. $\Delta/\sigma_{\rm R}$ is the ratio of the scatter after and before the FP fit and can be used to quantify the existence of the plane \citep{van_de_Sande_et_al.(2014)}.

Below, we first present in Section~\ref{sec:result1}, the plane rotation, then in Section~\ref{sec:result2}, the plane offset and the evolution of the scatter, and in Section~\ref{sec:result3}, the relation of the $M_{\rm dyn}/L - M_{\rm dyn}$ plane and its implication, where $M_{\rm dyn}$ denotes the dynamical mass. 

\begin{table*}
\begin{tabular}{cccccccc}
\hline
\hline
$z$ & $N_{\rm gal}$ & $a$ & $b$ & $c$ & $\Delta$ & $\frac{\Delta}{\sigma_{\mathrm{R}}}$ & $\delta$\\
\hline
0   & 1064 & $3.129\pm0.128$ & $1.295\pm0.026$ & $-0.627\pm0.012$ & $0.078\pm0.002$ & $0.287\pm0.010$ & $0.788\pm0.066$ \\
0.1 & 944 & $3.194\pm0.131$ & $1.290\pm0.027$ & $-0.630\pm0.011$ & $0.077\pm0.002$ & $0.282\pm0.010$ & $0.769\pm0.068$\\ 
0.2 & 852 & $3.087\pm0.134$ & $1.324\pm0.029$ & $-0.624\pm0.011$ & $0.078\pm0.002$ & $0.292\pm0.011$ & $0.829\pm0.065$\\ 
0.3 & 745 & $3.101\pm0.157$ & $1.303\pm0.036$ & $-0.617\pm0.012$ & $0.076\pm0.002$ & $0.288\pm0.011$ & $0.834\pm0.071$\\ 
0.4 & 665 & $3.023\pm0.155$ & $1.261\pm0.033$ & $-0.595\pm0.012$ & $0.069\pm0.004$ & $0.274\pm0.011$ & $0.880\pm0.070$\\ 
0.5 & 606 & $3.140\pm0.164$ & $1.241\pm0.035$ & $-0.602\pm0.013$ & $0.071\pm0.004$ & $0.281\pm0.012$ & $0.835\pm0.077$\\ 
0.7 & 513 & $3.194\pm0.167$ & $1.230\pm0.035$ & $-0.601\pm0.014$ & $0.077\pm0.004$ & $0.297\pm0.013$ & $0.826\pm0.077$\\ 
1.0 & 365 & $3.009\pm0.209$ & $1.218\pm0.041$ & $-0.572\pm0.017$ & $0.073\pm0.004$ & $0.280\pm0.016$ & $0.929\pm0.093$\\ 
1.5 & 200 & $3.758\pm0.200$ & $1.119\pm0.055$ & $-0.616\pm0.016$ & $0.075\pm0.004$ & $0.304\pm0.021$ & $0.656\pm0.086$\\ 
2.0 & 64 & $3.643\pm0.363$ & $0.994\pm0.093$ & $-0.569\pm0.034$ & $0.079\pm0.004$ & $0.405\pm0.048$ & $0.717\pm0.153$\\ \hline
\end{tabular}
\caption{The FP fit results for the selected ETG samples. $a$, $b$, and $c$ are defined in Eq.~(\ref{fpexpression}). $\Delta$ is the root-mean-square of the fitting residual $\mathrm{Res}$ ($\equiv a + b\log \sigma_{\mathrm{e}} + c \log I_{\rm e} - \log R_{\rm hsm}$). $\sigma_{\rm R}$ is the scatter in $\mathrm{log}\,R_{\rm hsm}$ (see Section~\ref{sec:results} for detailed definitions). $\delta$ is defined as $\delta\equiv 4c+b+2$. The errors of $a$, $b$, $c$, $\Delta$, $\Delta/\sigma_{\rm R}$, and $\delta$ are calculated with the bootstrap method.}
\vspace{2mm}
\label{table:table1}
\end{table*}

\subsection{The slope parameters $b$ and $c$ and their redshift evolution}
\label{sec:result1}
Observationally, rotation of the FP over a redshift range out to $z>1$ has been detected to a certain degree: $b$ with a local value of $\sim1.2$ (in $B$-band) decreases from $b\gtrsim 1$ at intermediate redshifts below $z\sim 0.8$ \citep[e.g.,][]{van_Dokkum_and_Franx(1996), Kelson_et_al.(2000), Wuyts_et_al.(2004), van_der_Marel_and_van_Dokkum(2007),Bolton_et_al.(2008),Auger_et_al.(2010),Saglia_et_al.(2010)} to $b \lesssim 0.8$ at $0.8 \lesssim z \lesssim 1.3$ \citep[e.g.,][]{di_Serego_Alighieri_et_al.(2005), Jorgensen_et_al.(2006)}. While $c$ from a local value of $\sim -0.8$ \citep[e.g.,][]{Jorgensen_et_al.(1996), Bender_et_al.(1998),Hou_and_Wang(2015)} increases to $\sim -0.6$ at redshifts $z\gtrsim 0.5$ \citep{di_Serego_Alighieri_et_al.(2005),Saglia_et_al.(2010)}. It is worth noting that despite this plane rotation, $b$ and $c$ evolve in a fashion such that Eq.~(\ref{eq:4cb2}) is broadly satisfied across the observed redshift ranges within measurement uncertainties\footnote{This still needs further confirmation using more extensive galaxy samples across cosmic time (e.g., see \citealt{Saglia_et_al.(2010)}).}.

\begin{figure}
\includegraphics[width=\columnwidth]{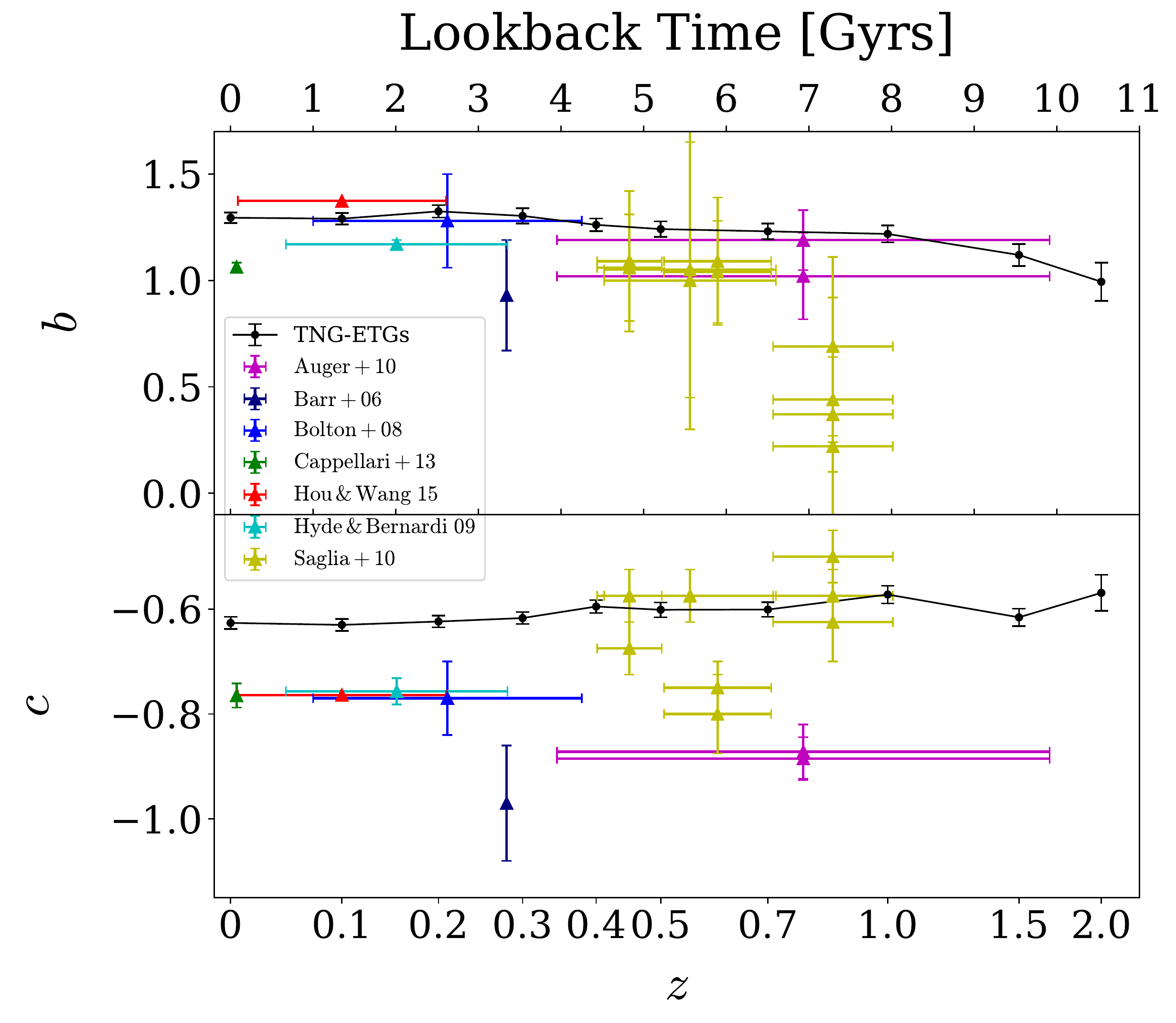}
\caption{Redshift dependencies of the FP slope parameters $b$ (upper) and $c$ (bottom). The results of the TNG100 ETG samples are indicated by the black circles with error bars calculated with the bootstrap method (see Section~\ref{sec:fpmethod}). Triangles with other colors show the observational results at different redshifts, with error bars indicating the uncertainties of the FP slope ($b$ and $c$) and the redshift range of their samples.}
\label{fig:bc}
\end{figure}

In comparison, the FP evolution of early-type galaxies from the TNG100 simulation demonstrates relatively mild plane rotation. As can be seen from Table~\ref{table:table1}, the slope parameters $b$ and $c$ vary only slightly with redshift. To better demonstrate this evolution, we present Fig.~\ref{fig:bc}, where the redshift evolution of the parameters is plotted against current measurements. We note here that only results where the investigated wavelength is close to the SDSS-$r$ band are included in the figure. Compared to observational data, the TNG predicted redshift evolution of $b$ agrees well with observations at redshifts below $z\lesssim 0.7$, but the magnitude of $b$ becomes marginally higher than observations at higher redshifts; the predicted evolution of $c$ is consistent with observations at $z\gtrsim0.4$, but appears systematically higher than the observed values at redshifts below $z \sim 0.4$. These inconsistencies may come from the larger sizes \citep{Genel_et_al.(2018),Rodriguez-Gomez_et_al.(2019)} and lower $\sigma_{\rm e}$ \citep{Wang_Y._et_al.(2020)} of IllustrisTNG ETGs compared with observations, as a result of overly-strong AGN feedback (typically the isotropic black hole kinetic winds in the AGN quiescent phase) in IllustrisTNG for puffing up the galaxies \citep{Wang_Y._et_al.(2019),Wang_Y._et_al.(2020)}.%However, we caution that the observational data also appear to have some inconsistencies among themselves. This might be caused by measurements at different wavelengths, or in different galaxy environments.

\subsection{The FP zero point $a$, its scatter and redshift evolution}
\label{sec:result2}
\begin{figure}
\centering
\includegraphics[width=\columnwidth]{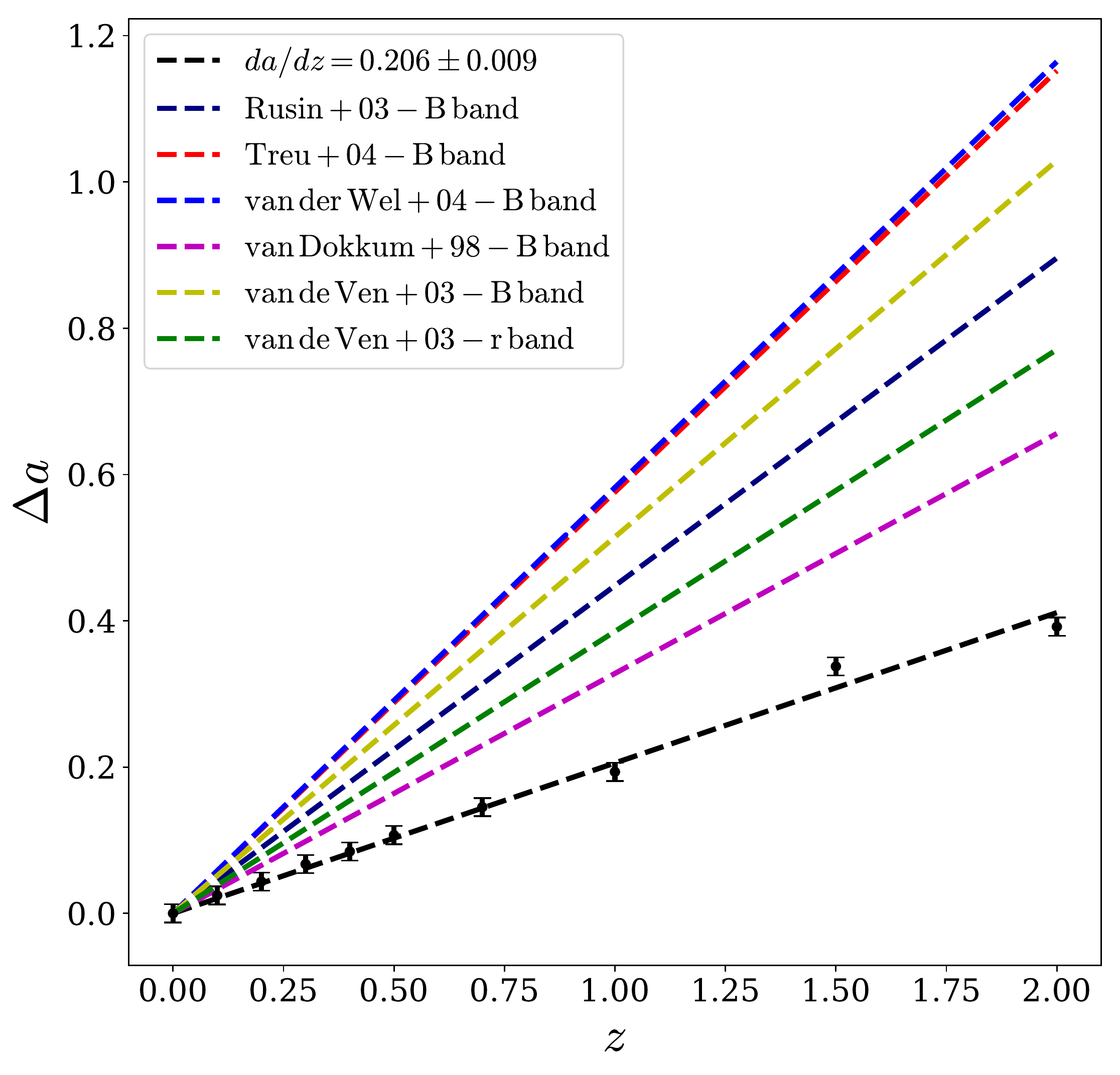}
\caption{Redshift dependencies of the relative zero point ($\Delta a\equiv a-a_{z=0}$). The error bars are calculated with the bootstrap method. The black dashed line is the best-fit line of the $\Delta\,a-z$ relation. The dashed lines with colors indicate the best-fit lines of the $\Delta\,a-z$ relation from observations listed in the legend.}
\label{fig:zp}
\end{figure}

The zero point of the observed FP changes dramatically with cosmic time. To demonstrate the redshift evolution of $a$, we present Fig.~\ref{fig:zp}, where the relative zero points ($\Delta a\equiv a-a_{z=0}$) at various redshifts are plotted. The relative zero points are obtained using fixed plane slopes $b=1.295$ and $c=-0.627$ (as they evolve mildly with redshift) from the $z=0$ plane fitting. The zero points of the TNG100 FP have changed by $\sim 0.1$, 0.2, and 0.4 out to $z=0.5$, 1.0, and 2.0, while observationally, the zero points have a steeper relation with redshift than TNG100 FPs \citep{Rusin_et_al.(2003),Treu_and_Koopmans(2004),van_der_Wel_et_al.(2004),van_Dokkum_et_al.(1998),van_de_Ven_et_al.(2003)}, including the results in the $r$-band \citep{van_de_Ven_et_al.(2003)}. We note here that the redshifts of observational samples are limited up to $z\sim 1.0$, but we see that the tight linear relation between $z$ and $\Delta a$ exists as early as $z\sim 2.0$ in TNG100.

Another interesting property of the FP is its tightness. It is related to questions such as how early a FP forms and how the tightness evolves with cosmic time. In order to answer these questions, we apply $\Delta$ to quantify the strength of the FP and $\Delta/\sigma_{\rm R}$ to quantify the existence of the FP (see Eq.~\ref{eq:scatter} for detailed definitions).

\begin{figure}
\includegraphics[width=\columnwidth]{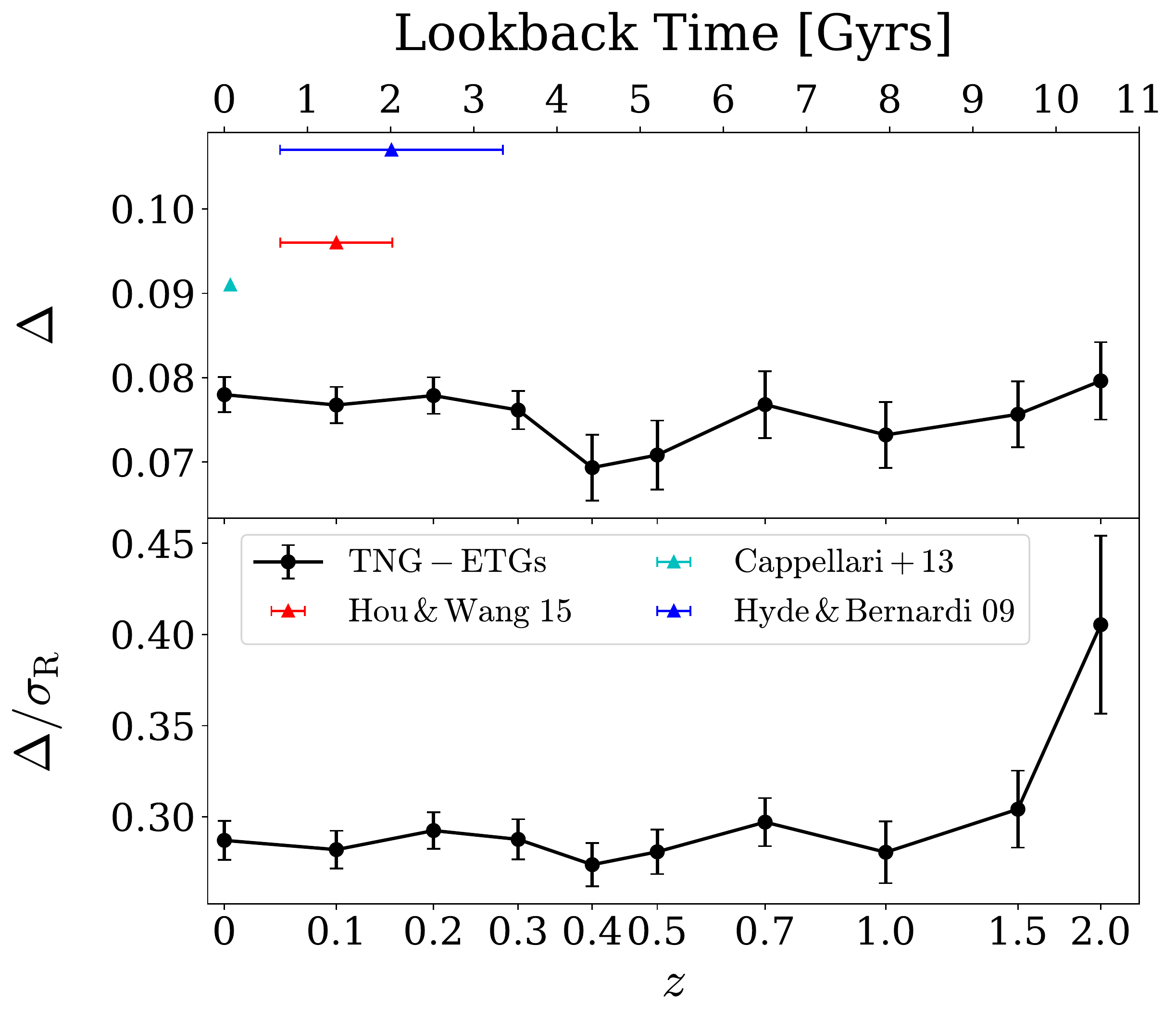}
\caption{The redshift dependence of $\Delta$ (upper) and $\Delta/\sigma_{\rm R}$ (bottom). The results of the TNG100 ETG samples are indicated by the black circles with error bars calculated with the bootstrap method. In the upper panel, triangles with colors are the results from observational works listed in the legend, with error bars indicating the redshift range of their samples.}
\label{fig:scatterratio}
\end{figure}

From Table~\ref{table:table1}, we have already seen that the FP exists as early as $z=2.0$ in the sense that its scatter $\Delta$ (and $\Delta/\sigma_{\rm R}$) is already small by then. We further present the redshift evolution for $\Delta$ and $\Delta/\sigma_{\rm R}$ in Fig.~\ref{fig:scatterratio}. As can be seen, the relative scatter $\Delta/\sigma_{\rm R}$ decreases from $\sim 0.4$ at $z=2.0$ to as low as $\sim 0.3$ at $z=1.5$, and stays nearly constant since then. The FP scatter, however, never undergoes an obvious variation across the various redshift ranges. Observationally, the observed scatter of the FP varies a lot from 0.091 \citep{Cappellari_et_al.(2013a)} to 0.107 \citep{Hyde_and_Bernardi(2009)}, showing an increasing trend towards high redshift. However, the FP scatter of simulated galaxies are typically lower than observed ones.

\begin{figure*} 
\includegraphics[width=\columnwidth]{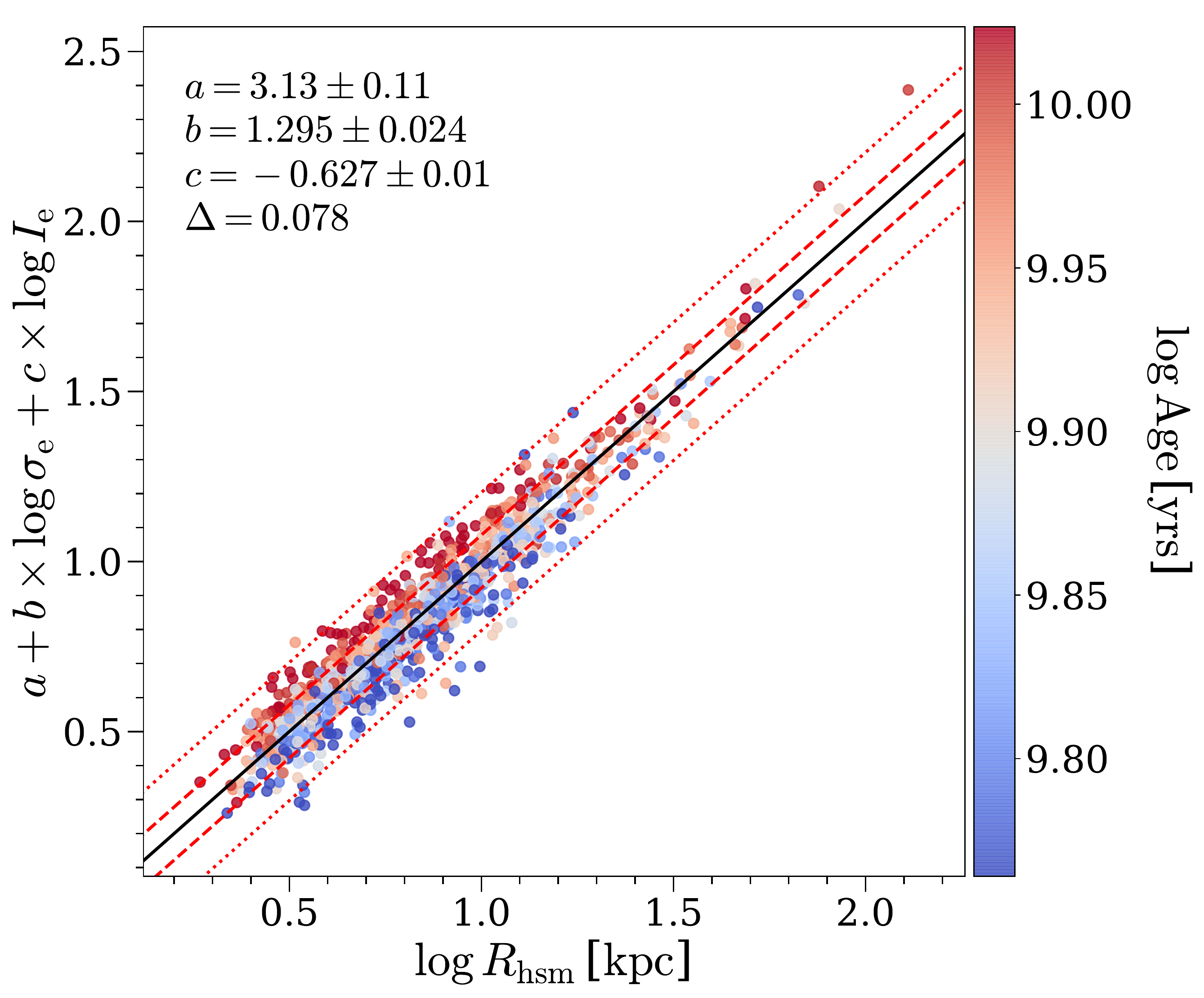}
\includegraphics[width=0.88\columnwidth]{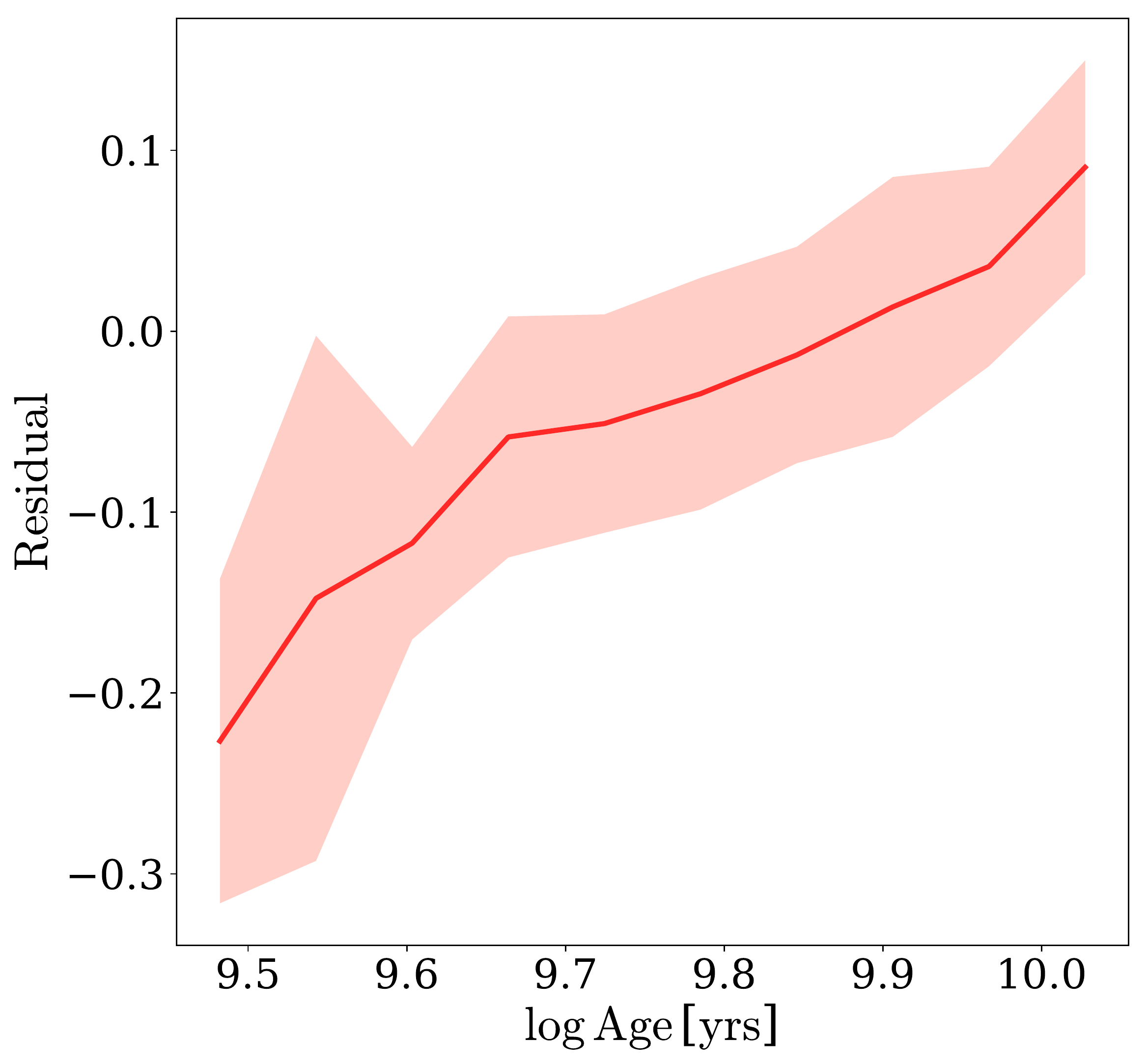}
\caption{Left panel: Edge-on view of the best-fit fundamental plane for early-type galaxies at $z=0$ in TNG100, color-coded by their stellar age, $\log \rm Age$. The black solid line is the best-fit plane and the red dashed lines show the 68\% ($1\sigma$) and 99\% confidence intervals. Right panel: Relation between the FP residual, $\rm Res$, and the stellar age, $\log \rm Age$. The FP residual is calculated as $\mathrm{Res}=a+b\times \mathrm{log}\,\sigma_{\rm e}+c\times \mathrm{log}\,I_{\rm e}-\mathrm{log}\,R_{\rm hsm}$. The red shaded region represents the $1\sigma$ range of residual.}
\label{fig:fp_paras}
\end{figure*}

The FP scatter has been identified to be linked to aging stellar populations in early-type galaxies \citep{van_Dokkum_and_Franx(1996), Forbes_et_al.(1998), Terlevich_and_Forbes(2002)}. In order to see such a connection, we present Fig.~\ref{fig:fp_paras}, where the left panel shows the $z=0$ FP color-coded by the ETGs' stellar age $\mathrm{log}\,\rm Age$, which is calculated as the luminosity-weighted age within an aperture of $R_{\rm hsm}$, and the right panel shows the correlation between the plane residual ($\rm Res$) and $\log \mathrm{Age}$. As can be seen, the FP residual ($\rm Res$) tightly correlates with age: it increases by $\sim 0.4\,\mathrm{dex}$ as $\log \mathrm{Age}$ changes from $\sim 9.5$ to $\sim 10$. This trend is also seen in observations \citep{Forbes_et_al.(1998),Terlevich_and_Forbes(2002),Graves_et_al.(2009),Springob_et_al.(2012),Magoulas_et_al.(2012)}. We have also investigated the correlations between the FP residual ($\rm Res$) and other properties of galaxies in TNG, i.e., metallicity and mass-to-light ratio and find that the correlations still exist. However, stellar age shows the strongest correlation with the FP residual, which is consistent with the observational results \citep{Magoulas_et_al.(2012)}. This indicates that our ETG sample selected from TNG100 supports the scenario where the galaxy stellar age is the fourth parameter of the FP relation \citep{Forbes_et_al.(1998)}.

\subsection{{\it A} plane versus {\it THE} plane: $\log M_{\rm dyn}/L - \log M_{\rm dyn}$ relation}
\label{sec:result3}
\begin{figure}
\centering
\includegraphics[width=\columnwidth]{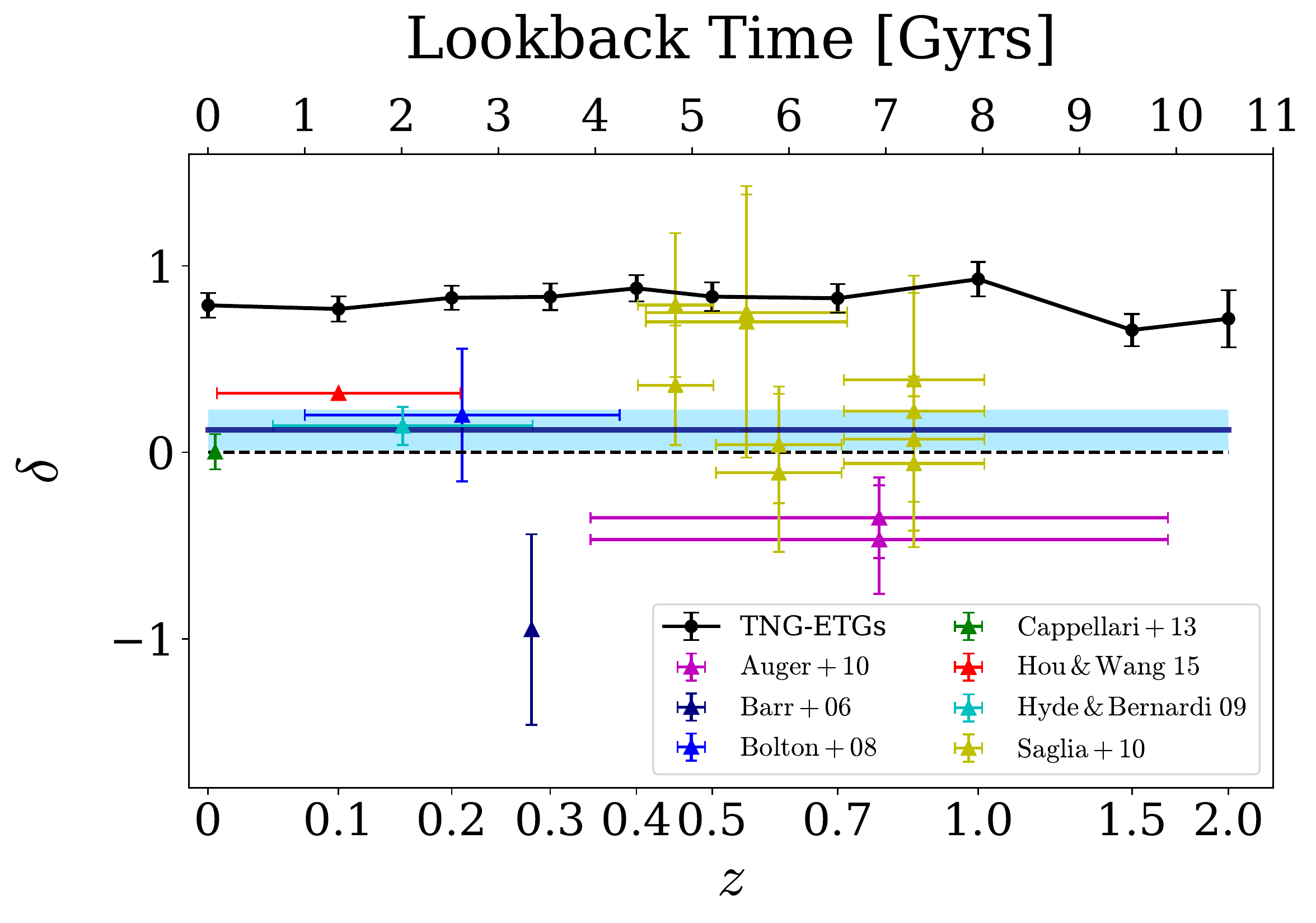}
\caption{Redshift dependence of $\delta\equiv 4c+b+2$. The results of TNG100 samples are indicated by the black circles and observational results are indicated by triangles with colors. For observational results, the errors of $\delta$ are calculated as $\sqrt{\sigma_{\rm b}^2+\left(4\sigma_{\rm c}\right)^2}$, where $\sigma_{\rm b}$ and $\sigma_{\rm c}$ are the plane fit errors. For TNG-ETGs, the errors of $\delta$ are calculated with the bootstrap method. The black dashed line represents $\delta=0$. The blue solid line represents the average value of $\delta$ for the observations, with the shaded region indicating the $1\sigma$ range of $\delta$.}
\label{fig:4cb}
\end{figure}

The seemingly mild inconsistencies between the simulated and observationally-constrained plane slopes $b$ and $c$ at various redshifts (as shown in Fig.~\ref{fig:bc}) actually point to an interesting theoretical question: how plausible is the plane relation seen in TNG100 compared to observations, given that the observed FPs till now satisfy Eq.~(\ref{eq:4cb2}) within their measurement uncertainties. To quantify this, we can define $\delta\equiv 4c+b+2$ at any given redshift and quantify how well the measurement of $\delta$ is clustered around zero, given the model uncertainties. For observational results, the measurement uncertainties of $\delta$ are calculated as $\sqrt{\sigma_{\rm b}^2+\left(4\sigma_{\rm c}\right)^2}$, with $\sigma_{\rm b}$ and $\sigma_{\rm c}$ being the measurement uncertainties of $b$ and $c$. For TNG FPs, the measurement uncertainties of $\delta$ are calculated with the bootstrap method. We present Fig.~\ref{fig:4cb} to show the values of $\delta$ at every redshift investigated in our work and compare them with observations \citep{Barr_et_al.(2006),Bolton_et_al.(2008),Hyde_and_Bernardi(2009),Auger_et_al.(2010),Saglia_et_al.(2010),Cappellari_et_al.(2013a),Hou_and_Wang(2015)}. As can be seen, $\delta$ of the FPs in TNG are typically larger than 0, which is inconsistent with the prediction of Eq.~(\ref{eq:4cb2}). Note that the errors of $b$ and $c$ for TNG FPs are somewhat underestimated because we did not take into account the measurement uncertainties of $\log R_{\rm hsm}$, $\log\,I_{\rm e}$, and $\log\,\sigma_{\rm e}$. However, even if the errors of $b$ and $c$ are increased by a factor of $\sim 3$, $\delta=0$ is still not satisfied. In contrast, the observed FPs' $\delta$ are clustered around 0 at various redshifts. The average observational value for $\delta$ is 0.120, with the $1\sigma$ range being 0.109, indicating that $\delta= 0$ is satisfied by the observed FPs at $1\sigma$ level. We note here that the $1\sigma$ range of observational $\delta$ is estimated as the intrinsic scatter, calculated as the square root of the difference between the variance of the observational $\delta$ and the average of their measurement uncertainties squared. 

\begin{figure}
\centering
\includegraphics[width=\columnwidth]{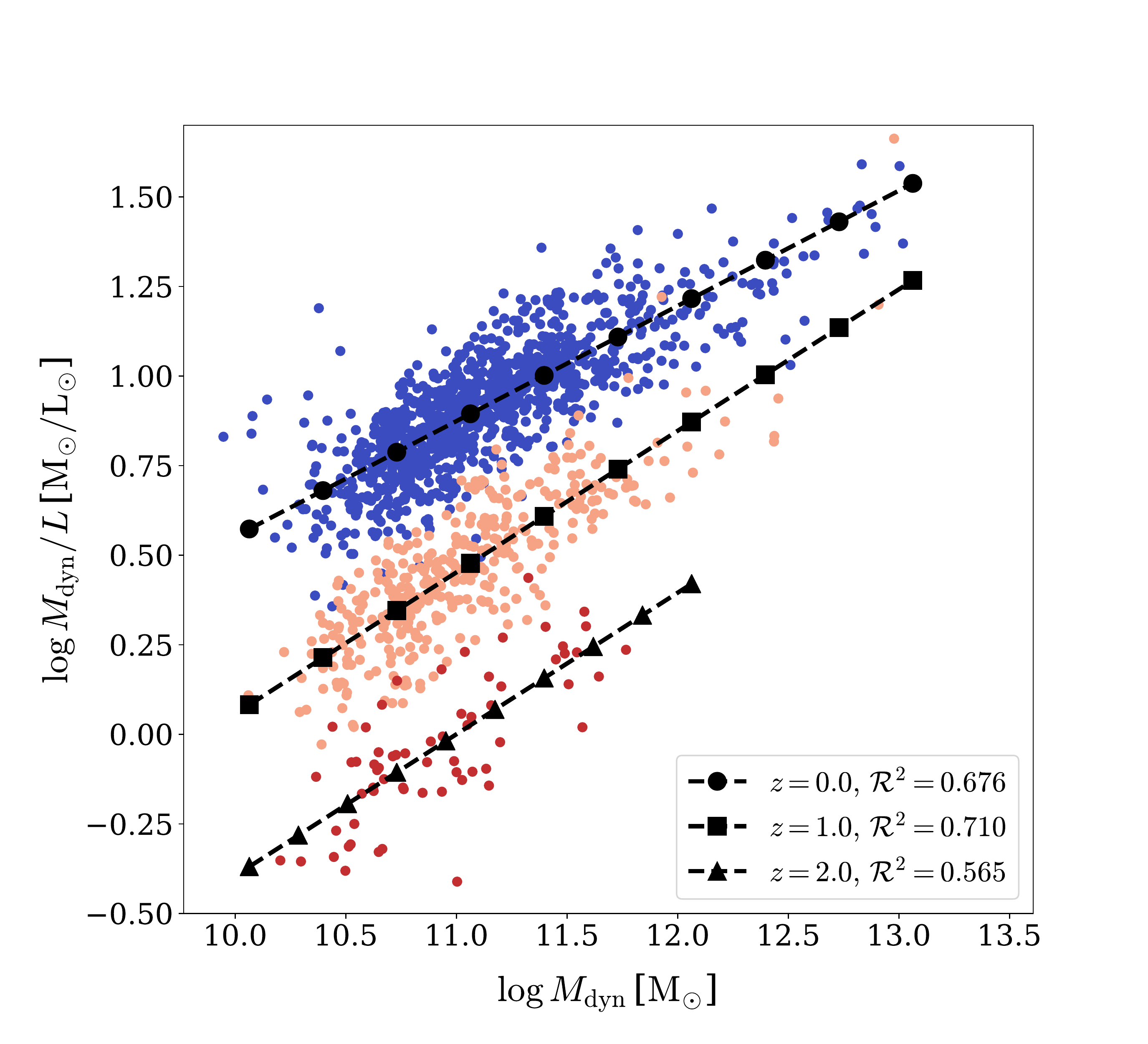}
\caption{Relation between the dynamical mass-to-light ratio ($\mathrm{log}\,M_{\rm dyn}/L$, where $M_{\rm dyn}\equiv 5\sigma_{\rm e}^2R_{\rm hsm}/G$) and the dynamical mass ($\mathrm{log}\,M_{\rm dyn}$) of galaxies at $z=0,1,$ and $2$. The dashed lines with different markers give the best linear-fit lines of the $\mathrm{log}\,M_{\rm dyn}/L$-$\mathrm{log}\,M_{\rm dyn}$ relation at each redshift. $\mathcal{R}^2$ is the coefficient of determination.}
\label{fig:ml2m}
\end{figure}

As already discussed in Section~\ref{sec:introduction}, $\delta = 0$ is equivalent to the fact that a certain dynamical mass-to-light ratio $\mathrm{log}\,M_{\rm dyn}/L$ (where $M_{\rm dyn}\equiv 5\sigma_{\rm e}^2R_{\rm hsm}/G$, \citealt{Bender_et_al.(1992)}) would follow a linear function of $\mathrm{log}\,M_{\rm dyn}$. To see how well such a relation is met by our simulated ETGs, we present Fig.~\ref{fig:ml2m}, where $\mathrm{log}\,M_{\rm dyn}/L$ versus $\mathrm{log}\,M_{\rm dyn}$ is plotted\footnote{Note that this is essentially the same as the $\sqrt{3}\kappa_3-\sqrt{2}\kappa_1$ projection of the FP relation (see \citealt{Bender_et_al.(1992)}).} for galaxies at $z=0,1,$ and 2 in our samples. A linear fit of $\mathrm{log}\,M_{\rm dyn}/L - \mathrm{log}\,M_{\rm dyn}$ is applied, with the coefficient of determination $\mathcal{R}^2$ indicating the goodness of linear fit. As can be seen, the  $\mathcal{R}^2$ of each fit is low ($\lesssim 0.7$), indicating that $\mathrm{log}\,M_{\rm dyn}/L$ does not have an acceptable linear relation with $\mathrm{log}\,M_{\rm dyn}$. In addition, observed galaxies have shown strong evolution of the slope of the $\mathrm{log}\,M_{\rm dyn}/L-\mathrm{log}\,M_{\rm dyn}$ relation (e.g., \citealt{Jorgensen_et_al.(2006),Saglia_et_al.(2010),Holden_et_al.(2005)}), which is, however, not seen for TNG100 early-type galaxies. Thus, the deviation from $\delta = 0$ for TNG100 ETGs arises from the combination of a loose $\mathrm{log}\,M_{\rm dyn}/L-\mathrm{log}\,M_{\rm dyn}$ linear relation and a lack of evolution of the slope of this linear relation, which indicates that the tight FP in TNG100 is ultimately not consistent with the observations.

\section{Conclusion and Discussion}
\label{sec:conclusion}
In this work, we have used early-type galaxies (ETGs) in the IllustrisTNG-100 simulation (TNG100) from $z=0$ to $z=2$ to study the evolution of the fundamental plane (the slopes, scatter, and zero point). We found that the FP slopes $b$ and $c$ vary mildly across the redshift range $z\in[0, 2]$. The slope parameter values at various redshifts are roughly consistent with observations within measurement uncertainties. In particular, for $b$, the result of TNG100 agrees with observations at $z \lesssim 0.7$, but is slightly higher at $z>0.7$. For $c$, the result of TNG100 is consistent with the observations at $z>0.4$, but is higher than the observational constraints at redshifts below $0.4$. 

We have also examined the relation between the relative zero point ($\Delta a\equiv a-a_{z=0}$) and the redshift for TNG100 FPs using fixed slope parameters $b=1.295$ and $c=-0.627$ of the FP at $z=0$. We find that the tight linear relation between $\Delta a$ and redshift found in observations is also seen in TNG100, but is somehow flatter than the observational results ($da/dz\sim 0.33-0.58$ in observations, but $0.206$ in TNG100). In observations, the results are limited to $z\sim 1.0$ due to the lack of high redshift samples, while in TNG100, galaxies at higher redshift ($z\sim 2.0$) are also found to obey this linear relation.

The scatter $\Delta$ of the FP in TNG100 stays as low as $\lesssim 0.8$ dex and is nearly unchanged since $z=2.0$, while $\Delta/\sigma_{\rm R}$ (see Section~\ref{sec:result2} for its definition) decreases from $\sim 0.4$ at $z=2.0$ to $\sim 0.3$ at $z=1.5$, and remains constant since then. This implies that the tight fundamental plane relation exists as early as $z=2.0$. In comparison, the FP scatter of TNG100 ETGs is typically lower than those in observations. Observationally, the FP scatter varies significantly from 0.091 to 0.107 across redshifts up to $z\sim 0.155$, with an increasing trend towards higher redshifts. 

To see the origin of the FP scatter, we have investigated where galaxies with different stellar ages are located on the edge-on view of the FP at $z=0$. We find that older galaxies sit near the `top' of the plane, and younger galaxies vice versa. The FP residual, which is calculated as $\mathrm{Res}\equiv a+b~ \mathrm{log}\,\sigma_{\rm e}+c~\mathrm{log}\,I_{\rm e}-\mathrm{log}\,R_{\rm hsm}$, strongly correlates with stellar age, indicating the fact that stellar age can be used as a crucial fourth parameter of the FP as pointed out by \citet{Forbes_et_al.(1998)}.

An interesting piece of observational evidence is that the observed FPs more or less satisfy Eq.~(\ref{eq:4cb2}). To quantify this, we define $\delta\equiv 4c+b+2$ and find that $\delta= 0$ is satisfied by observed FPs at $1\sigma$ level. We find that for TNG100 FPs, however, this condition is not satisfied. As can be seen in Fig.~\ref{fig:4cb}, the $\delta$ for TNG100 FPs is significantly inconsistent with $\delta=0$. This suggests that the plane relation in TNG100, despite being tight, is not an observationally plausible plane.

The inconsistency with $\delta=0$ indicates that the tight linear relation between the dynamical mass-to-light ratio $\mathrm{log}\,M_{\rm dyn}/L$ and the dynamical mass $\mathrm{log}\,M_{\rm dyn}$ (or equivalently the $\kappa_{3} - \kappa_{1}$ relation in \citealt{Bender_et_al.(1992)}) found in observations is not reproduced by the TNG100 simulation. This is confirmed in Fig.~\ref{fig:ml2m} where the $\mathrm{log}\,M_{\rm dyn}/L - \mathrm{log}\,M_{\rm dyn}$ relation has low coefficients of determination ($\mathcal{R}^2$). In addition, no redshift evolution of the slope for the $\mathrm{log}\,M_{\rm dyn}/L - \mathrm{log}\,M_{\rm dyn}$ relation is seen for the simulated galaxies, whereas the observed samples suggest a strong redshift evolution. 

Due to the diversification of the causes in shaping a galaxy, the $\mathrm{log}\,M_{\rm dyn}/L-
\mathrm{log}\,M_{\rm dyn}$ relation (or the $\delta=0$ condition) requires early-type galaxies to have the `correct' mixing between dark matter and baryons. In addition, halo assembly and galaxy formation are both shaped by various baryonic physical processes on smaller scales, as well as complicated merger and accretion environments on larger scales. Thus, the `correct' star-formation efficiencies that occur at a wide range of galaxy and halo mass scales are also required to form such a relation. In observations, $\delta=0$ is broadly satisfied, indicating that the observed early-type galaxies are well `tuned'. The simulated ETGs in TNG100, however, do not possess such properties. As an outlook, we suggest that in future simulations, tight power-law relations between $M_{\rm dyn}/L$ and $M_{\rm dyn}$ should be sought as some extra constraint to validate the simulations in terms of internal dynamical structure of galaxies in order to better reproduce galaxy populations which are observationally fully consistent. 

\section*{Acknowledgements}
We thank Dylan Nelson for helpful suggestions to this work. This work is partly supported by a joint grant between the DFG and NSFC (Grant No. 11761131004), the National Key Basic Research and Development Program of China (No. 2018YFA0404501), and grant 11761131004 of NSFC to SM.

%%%%%%%%%%%%%%%%%%%%%%%%%%%%%%%%%%%%%%%%%%%%%%%%%%

\bibliographystyle{mnras}
\bibliography{ref} 

\label{lastpage}
\end{document}